\documentclass[journal=jacsat,manuscript=article]{achemso}

\usepackage{amsmath,amssymb,bm}
\usepackage{enumitem}
\usepackage{xcolor}
\usepackage[version=3]{mhchem}
\usepackage[normalem]{ulem}
\usepackage[symbol]{footmisc}
\usepackage{siunitx}
\usepackage{physics}
\usepackage{booktabs}
\usepackage{textgreek}
\usepackage{hyperref}
\usepackage{booktabs}

\hypersetup{
    colorlinks=true,
    linkcolor=blue,
    citecolor=blue,
    urlcolor=blue
}

\SectionNumbersOn

\author{Ming-Zheng Du}
\affiliation{Department of Chemistry, Shanghai Key Laboratory of Molecular Catalysis and Innovative Materials, State Key Laboratory of Porous Materials for Separation and Conversion, Fudan University, Shanghai 200438, P. R. China}

\author{Shi-Yu He}
\affiliation{Department of Chemistry, Shanghai Key Laboratory of Molecular Catalysis and Innovative Materials, State Key Laboratory of Porous Materials for Separation and Conversion, Fudan University, Shanghai 200438, P. R. China}

\author{Jing Shen}
\affiliation{Department of Chemistry, Shanghai Key Laboratory of Molecular Catalysis and Innovative Materials, State Key Laboratory of Porous Materials for Separation and Conversion, Fudan University, Shanghai 200438, P. R. China}

\author{Ziyan Ye}
\affiliation{Department of Chemistry, Shanghai Key Laboratory of Molecular Catalysis and Innovative Materials, State Key Laboratory of Porous Materials for Separation and Conversion, Fudan University, Shanghai 200438, P. R. China}

\author{Jia-Xi Zeng}
\affiliation{
Dipartimento di Chimica, Universit\`{a} degli Studi di Milano, via Golgi 19, 20133 Milano, Italy}

\author{Dong H. Zhang}
\affiliation{State Key Laboratory of Chemical Reaction Dynamics, Dalian Institute of Chemical Physics, Chinese Academy of Sciences, Dalian 116023, P. R. China}
\affiliation{University of Chinese Academy of Sciences, Beijing 100049, China}

\author{Venkat Kapil}
\affiliation{Department of Physics and Astronomy, London Centre for Nanotechnology, Thomas Young Centre, University College London, WC1E 6BT London, United Kingdom}

\author{Wei Fang}
\affiliation{Department of Chemistry, Shanghai Key Laboratory of Molecular Catalysis and Innovative Materials, State Key Laboratory of Porous Materials for Separation and Conversion, Fudan University, Shanghai 200438, P. R. China}
\email{wei_fang@fudan.edu.cn}

\title{
\textbf{Isotope Effects at Classical Cost through Mass-Differentiable Machine Learning}\\
}

\begin{document}
\maketitle

\begin{abstract}
Isotope effects govern fractionation and modulate reactivity, with applications from hydrogen energy to environmental science and catalysis, yet predicting them requires resolving small isotope-dependent free-energy differences that remain very challenging for conventional path-integral simulations in complex systems.
Here we introduce iso-EPIGS, a path-integral coarse-graining framework built on a mass-differentiable neural network that reconstructs the mass- and temperature-dependent path integral centroid free-energy surface. 
Classical molecular dynamics on the learned surface yields rigorous isotope-resolved thermodynamics without explicit path-integral sampling. 
Benchmarks spanning gas, liquid, and crystalline phases, including liquid water and oxalic acid crystal, reproduce reference path-integral isotope free-energy differences, enthalpies, and lattice parameters at near classical computational cost.
Crucially, iso-EPIGS trained solely on all-H and all-D isotopologues retains high accuracy for unseen partially deuterated isotopologues, demonstrating robust transferability across nuclear masses.
Iso-EPIGS makes accurate isotope-effect simulations feasible for complex systems and lays the groundwork for foundation models of isotope effects across chemical space.
\end{abstract}

\section{Introduction}
Nuclear quantum effects (NQEs), including zero-point energy and tunneling, reshape molecular structure, thermodynamics, and reactivity.\cite{Markland_NatRevChem_2018_v2_p109,Ceriotti_ChemRev_2016_v116_p7529,Fang_IntRevPhysChem_2019_v38_p35} 
Because zero-point energies and tunneling probabilities depend strongly on nuclear mass, isotopic substitution can shift both chemical equilibria and reaction rates.\cite{BigeleisenMayer1947,Urey1947,WolfsbergVanHookPanethRebelo2009} 
Equilibrium isotope effects govern fractionation central to hydrogen-isotope separation\cite{WangChallaShollJohnson1999,WeinrauchEtAl2017,OhHirscher2016,liu_barely_2019} and environmental tracing,\cite{Craig1961,Dansgaard1964,Gat1996} whereas kinetic isotope effects—often amplified by tunneling—can strongly influence catalytic and enzymatic reactions.\cite{Kastner-2014-WIRE,masgrau_atomic_2006,KlinmanKohen2013,Schreiner2017} 
Predicting isotope effects in complex systems yet remains challenging in terms of both accuracy and computational cost. 
For isotope thermodynamics, fractionation and related observables are determined by differences between isotope-substitution free energies that can be very small, placing unusually stringent demands on accuracy.\cite{Ceriotti_JChemPhys_2013_v138_p14112,ZimmermannVanicek2009,ChengCeriotti2014} 
Although direct quantum calculations are feasible for small gas-phase systems,\cite{Barone2004,LiuTossellLiu2010,WebbMiller2014} rigorous treatments in condensed phases and other high-dimensional environments generally rely on path-integral simulations.\cite{feynman_quantum_1965,chandler_exploiting_1981,Markland_NatRevChem_2018_v2_p109,ChengBehlerCeriotti2016,BocusEtAl2023,GieseEtAl2022} 
Their many-bead representation multiplies the sampling cost, while separate simulations for different isotopologues compound it further.\cite{MarsalekEtAl2014} 
Together, these accuracy and cost barriers have limited predictive isotope thermodynamics in complex molecular systems.

An alternative to explicit ring-polymer sampling is offered by the centroid formulation, which integrates out fluctuations around the path-integral centroid and encodes their equilibrium effects in a quantum effective free-energy surface.\cite{feynman_quantum_1965,cao_formulation_1994,voth_path-integral_1996,Shi2003} 
Machine learning has made this formulation computationally practical: path-integral coarse-graining (PIGS) learns effective forces from path-integral data, enabling molecular dynamics on the learned surface at near-classical cost.\cite{musil_quantum_2022,loose_centroid_2022,Wu2022} 
By accelerating centroid molecular dynamics, these approaches have enabled efficient calculations of vibrational spectra and other approximate quantum dynamical properties in molecular and condensed-phase systems. 
However, force-based PIGS does not determine the centroid free energy itself or its temperature dependence, limiting its ability to predict quantitative quantum thermodynamics. 
We recently addressed these limitations with EPIGS (entropic path-integral coarse-graining), which reconstructs the centroid free energy and its temperature derivative to provide rigorous quantum thermodynamics across temperatures.\cite{EPIGS} 
Despite these advances, a unified description of isotope thermodynamics and kinetics remains an outstanding challenge. 
Both PIGS and EPIGS models could not learn the continuous mass dependence of quantum free energies.
Without this dependence, small isotope-dependent free-energy differences are difficult to resolve reliably.
Generalizing to unseen isotopologues compounds the challenge, as their combinatorial growth can demand substantially more path-integral reference data for training.

In this work, we extend EPIGS along the nuclear-mass dimension to develop isotopic EPIGS (iso-EPIGS). 
Nuclear masses enter the neural-network representation as continuous atomic descriptors, while centroid quantum kinetic energies provide extact labels for the mass derivatives of the centroid free energy. 
Combined with the free-energy, force, and entropy targets of EPIGS, this yields an effective surface differentiable with respect to coordinates, temperature, and individual nuclear masses, enabling isotope-substitution free energies to be evaluated through mass integration. 
We benchmark iso-EPIGS for the gas-phase formic acid dimer, liquid water, and crystalline oxalic acid, obtaining quantitative agreement with reference path-integral results. 
Notably, models trained only on fully hydrogenated and fully deuterated isotopologues accurately predict partially deuterated isotopologues absent from training. 
Together, these results demonstrate the capacity of mass-differentiable learning to deliver accurate and transferable isotope-effect simulations for complex systems at near classical cost.

\section{Methods}
In the imaginary-time path integral formulation of quantum statistical mechanics, the partition function is isomorphic to that of a classical ring polymer (RP),
which can be reformulated in terms of the RP centroid $\boldsymbol{x}_\mathrm{c}$, \cite{feynman_quantum_1965,voth_path-integral_1996}
\begin{equation}
    Z=\left(\prod_{i=1}^{N}\Lambda_i^{-3}\right)
\int \mathrm d\boldsymbol x_{\mathrm c}\,
\mathrm{e}^{-\beta A_{\mathrm c}(\boldsymbol x_{\mathrm c};\beta,\boldsymbol m)},
\end{equation}
where $\beta=1/k_\text{B}T$ is the inverse temperature, $N$ is the number of atoms, $\Lambda_i=\sqrt{2\pi\beta\hbar^2/m_i}$ is the thermal de Broglie wavelength of atom $i$, and $A_\text{c}$ is the centroid free energy, which encodes temperature- and mass-dependent quantum fluctuations around the centroid.
Its mass dependence captures the quantum free-energy contributions to equilibrium isotope effects.

\begin{figure}
    \centering
    \includegraphics[width=0.8\columnwidth]{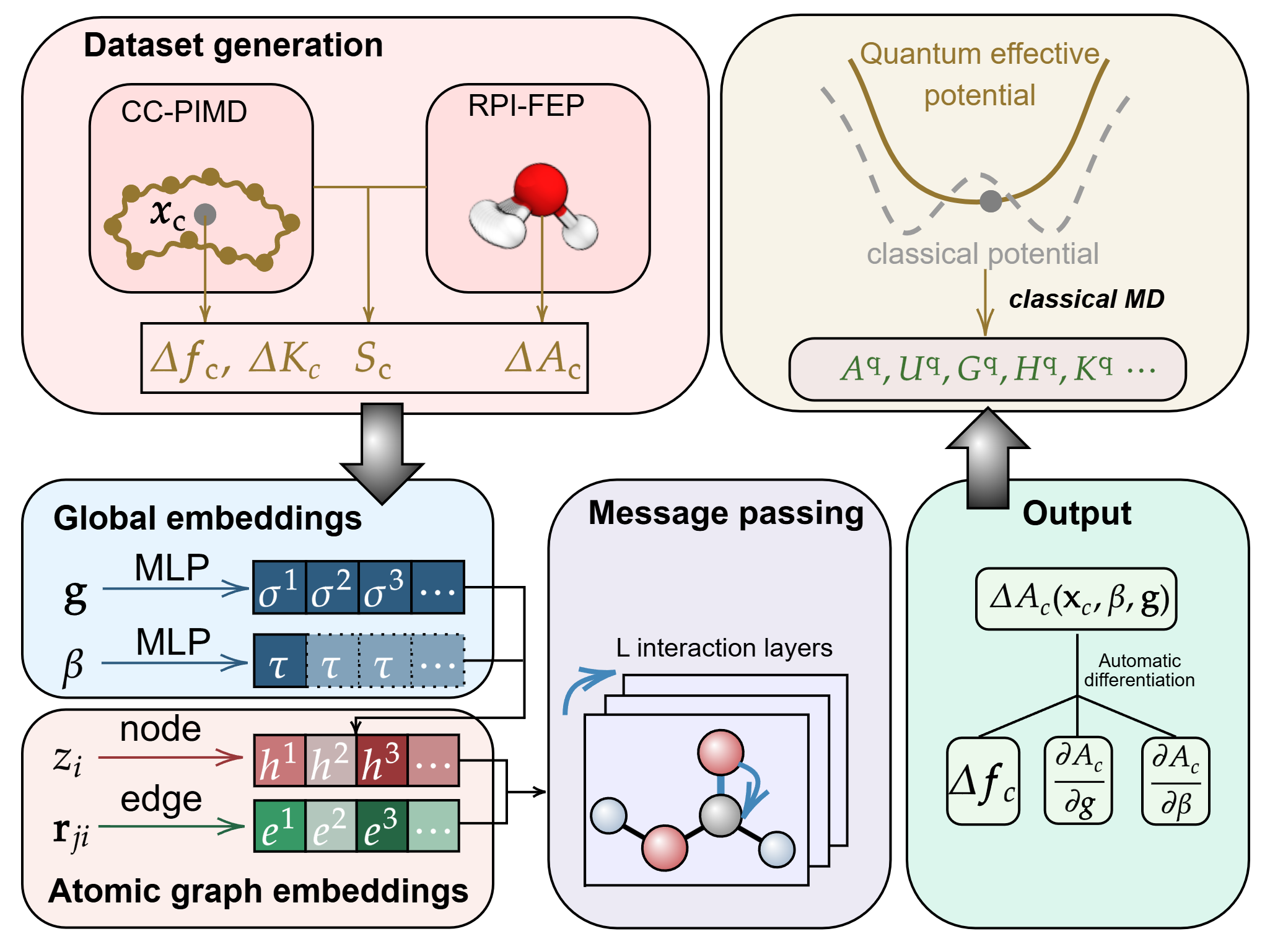}
    \caption{
Schematic of the iso-EPIGS framework.
Temperature ($\beta$) and mass ($\boldsymbol g$) embeddings augment the node features of an equivariant message-passing network, which is trained on fixed-centroid PIMD and RPI-FEP reference data to predict the centroid free-energy correction $\Delta A_{\mathrm c}(\boldsymbol x_{\mathrm c};\beta,\boldsymbol g)$.
Automatic differentiation of $\Delta A_{\mathrm c}$ yields $\Delta\boldsymbol f_{\mathrm c}$, $S_{\mathrm c}$ and $\Delta K_{\mathrm c,i}$, so that all four labels constrain a single thermodynamically consistent surface.
Classical MD on the resulting quantum effective potential then gives quantum thermodynamic properties, including isotope-substitution free energies, at classical sampling cost.
}
    \label{fig:framework}
\end{figure}

Iso-EPIGS reconstructs this temperature- and mass-dependent centroid free-energy surface using a neural network that takes the centroid coordinates, inverse temperature and individual nuclear masses as continuous inputs (Fig.~\ref{fig:framework}). 
Following the $\Delta$-learning strategy of EPIGS, the network predicts the quantum correction ($\Delta A_{\mathrm c}=A_{\mathrm c}-V(\boldsymbol x_{\mathrm c})$) to the underlying potential-energy surface $V$. 
Here we use a modified MACE equivariant graph neural network,\cite{NEURIPS2022_4a36c3c5} with atomic species and relative positions encoded in node and edge features, respectively, although the framework accommodates other graph neural-network potential architectures.
Two further inputs let a single model cover all temperatures and isotopic compositions: the inverse temperature $\beta$, which sets the extent of quantum delocalization, and the dimensionless mass descriptor $\boldsymbol g=\sqrt{\boldsymbol m^{*}/\boldsymbol m}$,\cite{Ceriotti_JChemPhys_2013_v138_p14112} evaluated elementwise, where $\boldsymbol m^{*}$ contains the masses of the most abundant isotopes of the corresponding elements. 
As a global thermodynamic variable, $\beta$ is mapped by a multilayer perceptron to an embedding that is shared by all atoms and augments the node features before message passing.
Nuclear masses are instead introduced locally, rather than broadcast to the whole system. 
This reflects the physics of isotopic substitution, whose effects depend on the substituted atom and its bonding environment.\cite{Polyakov2005GeometricMean}
We therefore keep the node features mass independent and introduce $\boldsymbol g$ only through an additional atomic readout,
\begin{equation}
\varepsilon_i=\sum_{n=1}^{3}c_n(\boldsymbol h_i,s_i)(g_i-1)^n,
\qquad
s_i=\sum_{j\in\mathcal B_i}w(r_{ij})(g_j-1),
\label{eq:isotope_readout}
\end{equation}
where $\boldsymbol h_i$ are the invariant features of atom $i$ from the embedding and message-passing layers, the coefficients $c_n$ are given by a multilayer perceptron, and $s_i$ sums the mass deviations of the atoms $\mathcal B_i$ covalently bonded to atom $i$, selected by a smooth switching function $w(r_{ij})$.
Each atom thus responds to its own mass through environment-dependent coefficients, while the bond term allows it to respond to the isotopic identities of its bonded neighbors.

The physical constraint on the learned mass dependence follows from the relation between free energy and nuclear kinetic energy. 
Nuclear masses enter the Hamiltonian only through the kinetic-energy terms ($\hat{\boldsymbol p}_i^{2}/2m_i$), giving the exact thermodynamic relation\cite{Ceriotti_JChemPhys_2013_v138_p14112}
\begin{equation}
\Delta K_{\mathrm c,i}
\equiv K_{\mathrm c,i}-\frac{3}{2\beta}
=-m_i\frac{\partial \Delta A_{\mathrm c}}{\partial m_i}
=\frac{g_i}{2}\frac{\partial\Delta A_{\mathrm c}}{\partial g_i},
\label{eq:Kc}
\end{equation}
where $K_{\mathrm c,i}$ is the mean quantum kinetic energy of atom $i$ at fixed centroids. 
The classical contribution ($3/(2\beta)$) is accounted for by the thermal-wavelength prefactor in the partition function. 
Quantum kinetic energies thus provide direct supervision of the response of the learned surface to each nuclear mass. 
Together with the force correction ($\Delta\boldsymbol f_{\mathrm c}=-\nabla_{\boldsymbol x_{\mathrm c}}\Delta A_{\mathrm c}$) and centroid entropy ($S_{\mathrm c}=k_{\mathrm B}\beta^2\partial\Delta A_{\mathrm c}/\partial\beta$),\cite{EPIGS} these quantities are obtained by automatic differentiation of the network output (Fig.~\ref{fig:framework}), linking all predictions through a single thermodynamically consistent surface.

To incorporate these physical constraints into training, we construct reference data at selected centroid configurations, temperatures and nuclear masses. 
Centroid-constrained PIMD supplies the force and atomic kinetic-energy corrections, while the ring-polymer instanton free-energy perturbation (RPI-FEP) approach\cite{EPIGS} provides the centroid free-energy correction. 
The entropy is obtained by combining the centroid free energy with the corresponding constrained mean energy. 
Each dataset entry therefore contains the inputs $(\boldsymbol x_{\mathrm c},\beta,\boldsymbol g)$ and four labels: $\{\Delta A_{\mathrm c}, \Delta\boldsymbol f_{\mathrm c}, S_{\mathrm c}, (\Delta K_{\mathrm c,i})\}$. 
The model is optimized using the joint loss 
\begin{equation}
\mathcal L=w_A\mathcal L_A+w_f\mathcal L_f+w_S\mathcal L_S+w_K\mathcal L_K, 
\end{equation}
where the four terms measure errors against the corresponding reference labels and the weights balance their contributions. 

Once trained, iso-EPIGS predicts the centroid free-energy correction at continuously varying temperatures and isotopic compositions. 
Automatic differentiation of the network output yields the force correction, centroid entropy and atomic kinetic-energy corrections (Fig.~\ref{fig:framework}). 
Classical MD on the reconstructed surface provides the centroid ensemble averages required to evaluate quantum thermodynamic properties. 
Because iso-EPIGS provides both $\Delta A_{\mathrm c}$ and its mass derivatives, free-energy differences can be obtained through two thermodynamic-integration (TI) routes.
In EPIGS-TI,\cite{EPIGS} the correction is switched on at fixed masses through a coupling parameter $\lambda$,
\begin{equation}
\Delta A^{\mathrm{q}\leftarrow\mathrm{cl}}\equiv A^{\mathrm q}-A^{\mathrm{cl}}=\int_0^1\langle\Delta A_{\mathrm c}\rangle_{\lambda}\,\mathrm d\lambda,
\label{eq:epigs_ti}
\end{equation}
where $\langle\cdot\rangle_{\lambda}$ is a classical average on $V+\lambda\Delta A_{\mathrm c}$.
In Mass-TI, $\lambda=1$ is fixed and the masses of the substituted atoms $\mathcal S$ are varied through a common $g$ from $g=1$ (H) to $g_{\mathrm D}=\sqrt{m_{\mathrm H}/m_{\mathrm D}}$,
\begin{equation}
\Delta A^{\mathrm{H}\rightarrow\mathrm{D}}=-\int_{g_{\mathrm D}}^{1}\sum_{i\in\mathcal S}\left\langle\frac{\partial\Delta A_{\mathrm c}}{\partial g_i}\right\rangle_{g}\mathrm dg,
\label{eq:mass_ti}
\end{equation}
where the classical free-particle term is omitted because it cancels between environments with the same substituted atoms.
Mass-TI is the conventional PIMD route,\cite{Ceriotti_JChemPhys_2013_v138_p14112,Rossi2015,Fang_JPhysChemLett_2016_v7_p2125} whereas iso-EPIGS evaluates its integrand from the learned derivatives (Eq.~\ref{eq:Kc}) under classical sampling.
Since classical free energies depend on mass only through trivial prefactors, the two routes satisfy $\Delta A^{\mathrm{H}\rightarrow\mathrm{D}}=\Delta A^{\mathrm{q}\leftarrow\mathrm{cl}}(\mathrm D)-\Delta A^{\mathrm{q}\leftarrow\mathrm{cl}}(\mathrm H)$, which we use as an internal consistency check (Section~\ref{sec:thermo}).

\section{Results}

We apply iso-EPIGS to three representative hydrogen-bonded systems spanning gas, liquid, and solid phases: the formic acid dimer, liquid water, and crystalline oxalic acid. 
In all three systems, the underlying potential $V$ is the medium MACE-OFF24 model,\cite{Kovacs_JAmChemSoc_2025_v147_p17598} which we fine-tuned\cite{kaur_data-efficient_2024} for crystalline oxalic acid and the formate anion with reference data at the level of theory of the original model (SI).
Across these systems, we first benchmark the learned centroid free-energy surfaces and their mass derivatives against path-integral references for individual configurations. 
We then explore the physical consequences of isotopic substitution, progressing from dimer binding and dissociation free energies to the quantum kinetic energy and enthalpy of liquid water, and finally to the enthalpy and lattice parameters of crystalline oxalic acid. 
Throughout, we test whether models trained only on fully hydrogenated and fully deuterated systems can predict isotope effects in partially deuterated isotopologues that are absent from training.

\subsection{Accurate learning of the centroid surface}

We train separate iso-EPIGS models for the formic acid, water, and oxalic acid systems, combining monomer and corresponding cluster or crystal configurations in the training set for each system.
Computational details of the fixed-centroid PIMD and RPI-FEP calculations used to generate the training data are provided in the Supplementary Information (SI), together with parity plots of training, validation, and test errors.
The RMSEs in $A_{\mathrm c}$ and $TS_{\mathrm c}$ remain below $0.5\,\mathrm{meV/atom}$, while the relative RMSE in $\boldsymbol f_{\mathrm c}$ is below $5\%$, showing that mass conditioning preserves the accuracy of EPIGS models trained on the same configurations.
The centroid-constrained quantum kinetic energy ($K_{\mathrm c}$) is also accurately reproduced, with RMSEs below $1.5\,\mathrm{meV/atom}$ in every system. 
Together, these results demonstrate that iso-EPIGS accurately captures the coordinate, temperature and nuclear-mass dependence of the centroid free-energy surface across molecular, liquid and solid environments.

\subsection{Isotopic thermodynamics at classical cost}

\subsubsection*{Gas-phase isotope effects in binding and dissociation}

Configuration-level accuracy does not by itself guarantee accurate isotope thermodynamics, which depends on small differences between ensemble-averaged free energies.
We therefore use the formic acid dimer, held together by two cooperative O--H$\cdots$O hydrogen bonds, as a representative gas-phase test.\cite{Li_ProcNatlAcadSciUSA_2011_v108_p6369,Fang_JPhysChemLett_2016_v7_p2125,EPIGS}
We quantify isotope effects on binding through the quantum contribution to the binding free energy, $\Delta\Delta A_{\mathrm{bind}}^{\mathrm{q}\leftarrow\mathrm{cl}}=\Delta A_{\mathrm{dimer}}^{\mathrm{q}\leftarrow\mathrm{cl}}-2\Delta A_{\mathrm{monomer}}^{\mathrm{q}\leftarrow\mathrm{cl}}$, where $\Delta A^{\mathrm{q}\leftarrow\mathrm{cl}}$ is the free-energy change upon quantizing the nuclei.
Because classical binding free energies do not depend on nuclear mass, the change of this quantity upon deuteration is the full isotope effect on binding.
Resolving them requires sub-meV precision in differences between species-level quantum corrections of several hundred meV.
For HCOOH and DCOOD, iso-EPIGS reproduces the species-level corrections to within 1.2~meV, and the correlated deviations of monomers and dimers largely cancel in $\Delta\Delta A_{\mathrm{bind}}^{\mathrm{q}\leftarrow\mathrm{cl}}$, which agrees with PIMD to within 0.2~meV for both dimers.
Deuteration makes binding more favorable: the quantum contribution decreases by $1.25\pm0.07$~meV from $(\mathrm{HCOOH})_2$ to $(\mathrm{DCOOD})_2$ in iso-EPIGS, compared with $1.3\pm0.4$~meV in mass-TI.
Because this reference uncertainty is comparable to the smallest shift, the comparison is limited by the precision of mass-TI rather than by iso-EPIGS ($\pm$0.13~meV).

Because iso-EPIGS supports both thermodynamic-integration routes, the deuteration free energy of each species can be computed twice, which provides a test that requires no reference data.
EPIGS-TI obtains it from the difference of the quantum contributions of DCOOD and HCOOH, whereas Mass-TI integrates the learned mass derivatives from $g=1$ to $g_{\mathrm D}$ (Eq.~\ref{eq:mass_ti}), so that together they form a closed thermodynamic cycle.
The cycle closes to 0.1~meV or better for the monomer, the dimer, and the binding shift, so sampling and quadrature errors are negligible and larger deviations from PIMD reflect the learned surface itself.
By contrast, the PIMD cycle closes only to within 0.35~meV, because the all-mass Mass-TI integrand varies steeply near the classical limit (SI).

Hydroxyl deuteration also changes gas-phase acidity, which we probe through the dissociation reactions $\mathrm{HCOOH}\rightarrow\mathrm{HCOO^-}+\mathrm{H^+}$ and $\mathrm{HCOOD}\rightarrow\mathrm{HCOO^-}+\mathrm{D^+}$.
Iso-EPIGS predicts a difference between the deuterated and hydrogenated dissociation free energies of 68.18 meV, compared with 67.37 meV from reference path-integral calculations.
Because both reactions produce the same formate anion, its free energy cancels from this difference, directly connecting the isotope shift to the relative free energies of the neutral acids and the released nuclei.
Together, the binding and dissociation results demonstrate that iso-EPIGS resolves isotope effects on both intermolecular association and acid dissociation within a common mass-dependent representation.

\subsubsection*{Isotope thermodynamics and transferability in liquid water}

Liquid water poses a different challenge: nuclear quantum fluctuations are coupled through an extended hydrogen-bond network, and their effects change with temperature and phase.\cite{Ceriotti_ChemRev_2016_v116_p7529,Chen_PhysRevLett_2003_v91_p215503,Morrone_PhysRevLett_2008_v101_p17801,ceriotti_nuclear_2013}
A single iso-EPIGS model, trained exclusively on $\mathrm{H_2O}$ and $\mathrm{D_2O}$, is applied to both gas and liquid phases over 281–409 K.
For these two isotopic endpoints, iso-EPIGS closely reproduces the PIMD quantum kinetic-energy correction, $K^{\mathrm q}-K^{\mathrm{cl}}$, throughout the temperature range (Fig. 2, upper panels).
The correction is substantially smaller for $\mathrm{D_2O}$ and decreases with increasing temperature in both phases.\cite{Ceriotti_PhysRevLett_2012_v109_p100604,wang_quantum_2014}
Beyond these overall trends, iso-EPIGS resolves the smaller difference between liquid and gas, with the liquid consistently exhibiting a larger kinetic-energy correction.\cite{MarklandBerne2012,ChengBehlerCeriotti2016}
Quantum potential energies also agree closely with PIMD (Supplementary Information), providing an accurate description of both contributions to the internal energy.

These energy contributions yield vaporization enthalpies in close agreement with PIMD for both isotopologues (Fig. 2, lower panels). NQEs reduce $\Delta H_{\mathrm{vap}}$ relative to classical MD, reflecting a net weakening of the liquid’s energetic cohesion. Zero-point motion weakens the hydrogen-bond network in water, reducing the energy required for vaporization.\cite{Li_ProcNatlAcadSciUSA_2011_v108_p6369,habershon_competing_2009,EPIGS} Deuteration suppresses this quantum contribution, resulting in the higher vaporization enthalpy of $\mathrm{D_2O}$,\cite{JancsoVanHook1974,HerrigEtAl2018} which is in between $\mathrm{H_2O}$ and classical water. Iso-EPIGS captures both this isotope dependence and the weakening of the quantum correction with increasing temperature.

\begin{figure}
    \centering
    \includegraphics[width=0.8\textwidth]{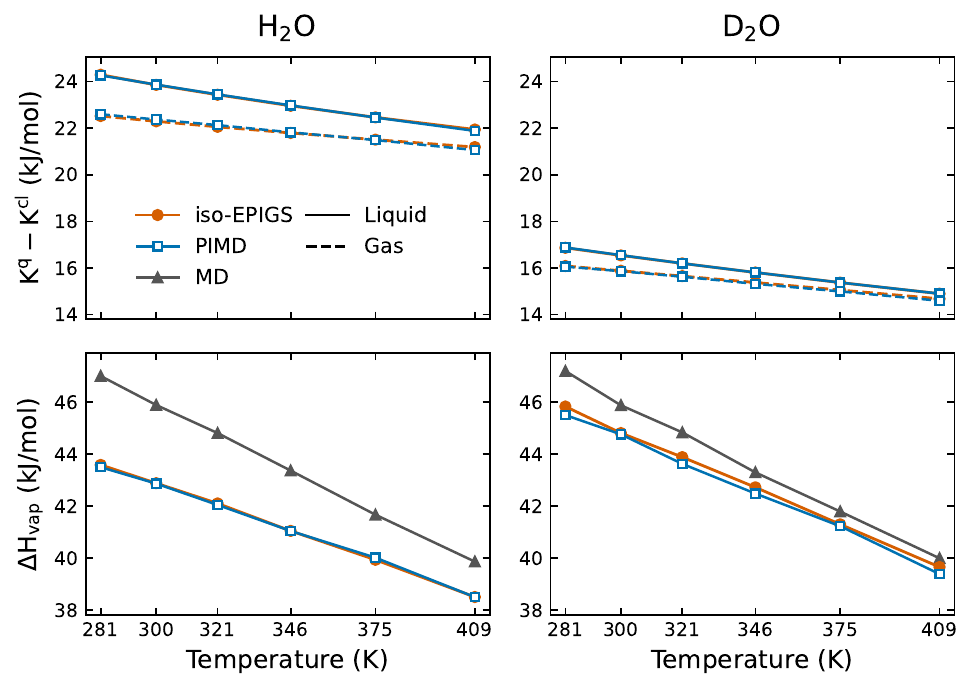}
\caption{%
Temperature-, phase-, and isotope-dependent thermodynamics of water.
The top row shows the excess of the quantum over the classical kinetic energy per molecule, $K^{\mathrm{q}}-K^{\mathrm{cl}}$, in the liquid (solid lines) and gas (dashed lines); the bottom row gives the enthalpy of vaporization, $\Delta H_{\mathrm{vap}}$.
Columns correspond to H$_2$O and D$_2$O (left to right), each simulated at six temperatures from 281 to 409~K, evenly spaced in $1/T$.
Symbols denote iso-EPIGS (filled orange circles), reference PIMD (open blue squares), and classical MD (gray triangles, bottom row only).
Only H$_2$O and D$_2$O were included in the iso-EPIGS training data.
Error bars are omitted because the statistical uncertainties are small compared with the plotted values.
Lines are guides to the eye.
}
    \label{fig:water_K_hvap}
\end{figure}

\subsubsection*{Solid-state isotope effects in crystalline oxalic acid}

Extending these benchmarks to the solid state, we assess iso-EPIGS through the sublimation thermodynamics and equilibrium lattice structure of crystalline $\beta$-oxalic acid.\cite{DerissenSmit1974,Rossi_PhysRevLett_2016_v117_p115702,Kapil_ProcNatlAcadSciUSA_2022_v119_pe2111769119} At 300 K, NQEs raise $\Delta H_{\mathrm{sub}}$ by 1.0 and 0.5 kJ/mol relative to classical MD for the fully hydrogenated and deuterated crystals, respectively (Table 1). Iso-EPIGS predicts $87.90\pm0.08$ and $87.52\pm0.09$ kJ/mol, agreeing with the corresponding PIMD values of $87.79\pm0.10$ and $87.41\pm0.11$ kJ/mol. By reproducing these increases, iso-EPIGS captures the enhanced crystal cohesion associated with quantum strengthening of the hydrogen-bond network. The model thus captures both quantum weakening and strengthening of intermolecular cohesion across diverse hydrogen-bond environments.\cite{Li_ProcNatlAcadSciUSA_2011_v108_p6369}
Beyond these energetic changes, iso-EPIGS also reproduces the temperature and isotope dependence of the lattice parameter $a$, parallel to the hydrogen-bond direction (Fig. 3). 
Across 250–375 K, it captures the PIMD thermal expansion and the larger quantum contraction in the hydrogenated crystal, giving $a_{\mathrm H}<a_{\mathrm D}<a_{\mathrm{cl}}$.
The expansion upon deuteration corresponds to the normal Ubbelohde effect, in which deuteration lengthens hydrogen bonds.\cite{RobertsonUbbelohde1939,UbbelohdeGallagher1955}
Although iso-EPIGS slightly underestimates the absolute lattice parameter, it closely reproduces the separation between the quantum H and D curves.
The contraction along the hydrogen-bond direction is consistent with the strengthening inferred from the sublimation enthalpy.
Agreement in both observables demonstrates that iso-EPIGS captures the influence of nuclear mass on crystal energetics and equilibrium structure within the same learned free-energy representation.

\begin{figure}
\centering
\includegraphics[width=0.8\columnwidth]{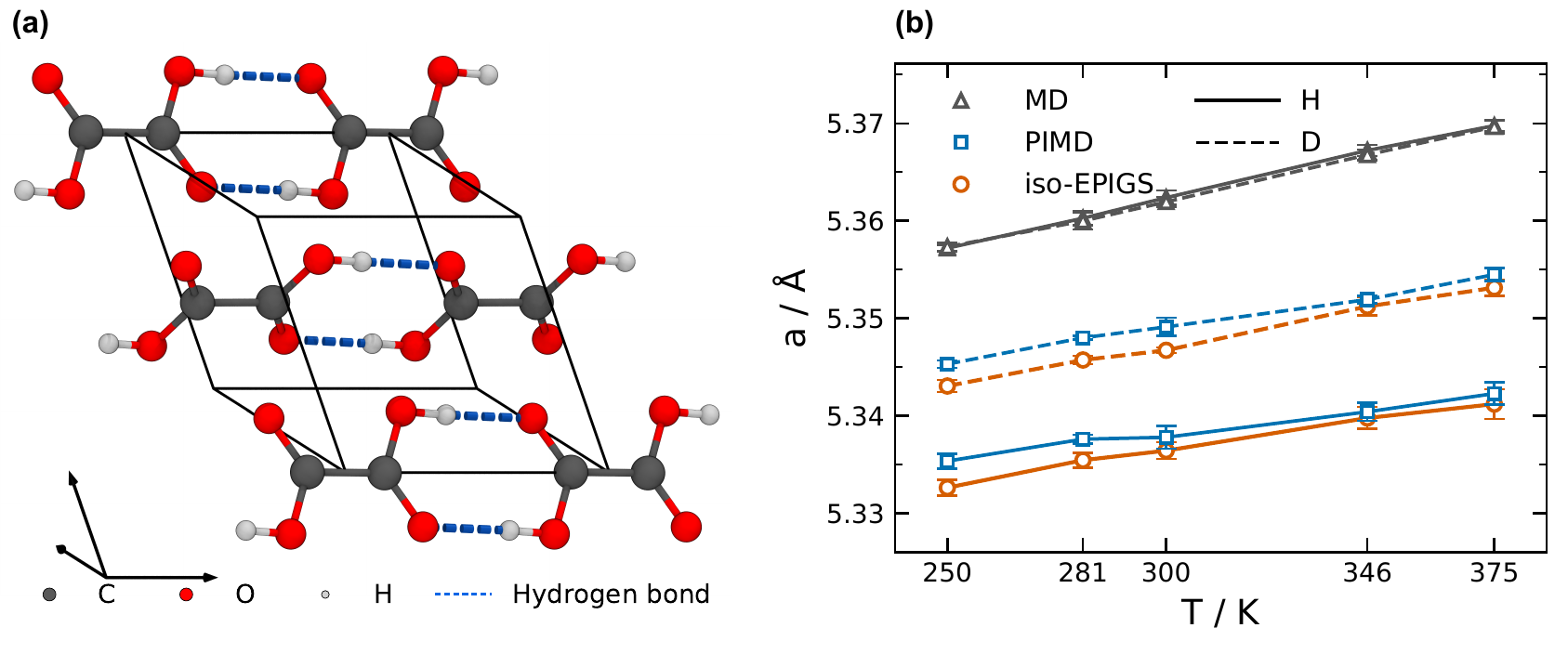}
\caption{Temperature dependence of the lattice parameter $a$ of $\beta$-oxalic acid, the axis parallel to the hydrogen-bond direction. Results are shown for fully protiated $\mathrm{H_2C_2O_4}$ (H, solid lines) and fully deuterated $\mathrm{D_2C_2O_4}$ (D, dashed lines) from classical MD (triangles), PIMD (squares), and iso-EPIGS (circles). The classical MD curves for the two isotopologues coincide within statistical uncertainty, reflecting the mass independence of classical structural properties. Error bars denote standard deviation over 5 independent runs.}
\label{fig:oxalic}
\end{figure}

\begin{table}[t]
    \centering
    \caption{
        Sublimation enthalpy, $\Delta H_{\mathrm{sub}}$, of $\beta$-oxalic acid at 300~K from classical MD, PIMD, and iso-EPIGS.
        Uncertainties are standard deviations over 7 independent runs.
    }
    \label{tab:oxalic_sublimation_enthalpy}
    \begin{tabular}{lccc}
        \toprule
        & \multicolumn{1}{c}{MD (kJ/mol)}
        & \multicolumn{1}{c}{PIMD (kJ/mol)}
        & \multicolumn{1}{c}{iso-EPIGS (kJ/mol)} \\
        \midrule
        (COOH)$_2$ & $86.81 \pm 0.08$ & $87.79 \pm 0.10$ & $87.90 \pm 0.08$ \\
        (COOD)$_2$ & $86.94 \pm 0.09$ & $87.41 \pm 0.11$ & $87.52 \pm 0.09$ \\
        \bottomrule
    \end{tabular}
\end{table}

\subsection{Mixed isotopologues: HDO and isotope fractionation}
\label{sec:hod}
\begin{figure}[htbp]
  \centering
  \includegraphics[width=0.8\linewidth]{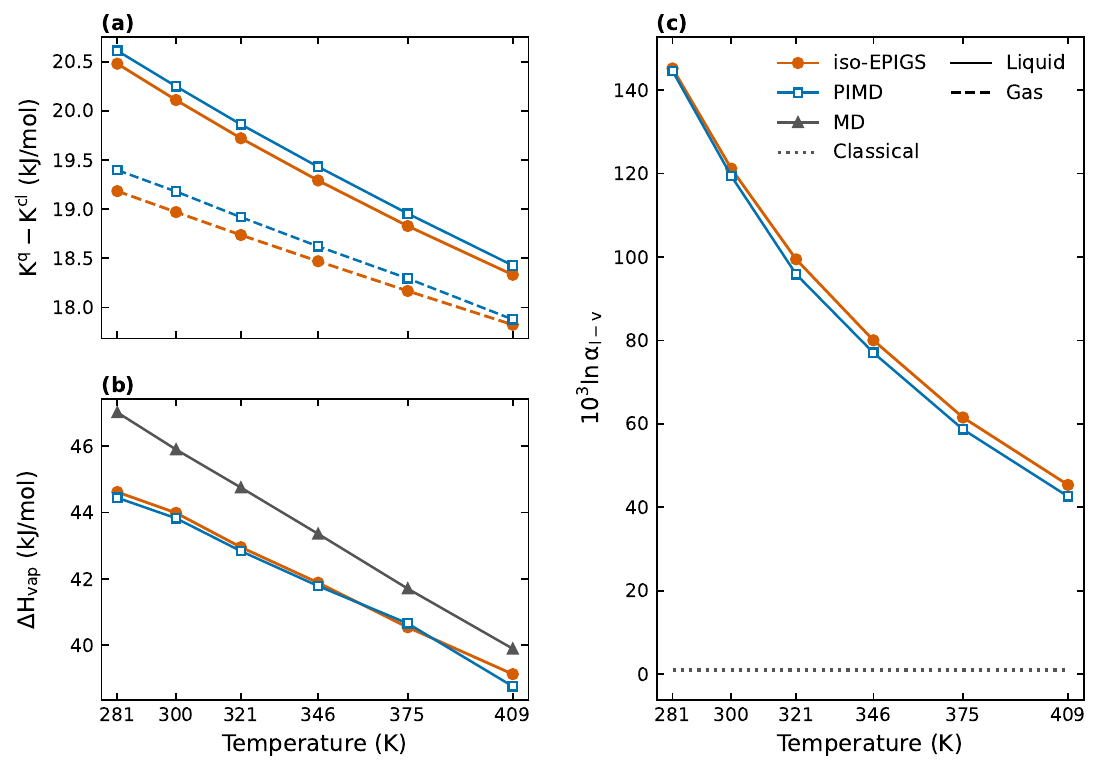}
    \caption{%
    Temperature dependence of the kinetic energy, vaporization enthalpy, and liquid--vapor fractionation of HDO.
    (a) Difference between the kinetic energy from the quantum simulations, $K^\mathrm{q}$, and that from classical MD, $K^\mathrm{cl}$, in the liquid (solid lines) and gas (dashed lines) phases.
    (b) Enthalpy of vaporization, $\Delta H_\mathrm{vap}$.
    (c) Liquid--vapor fractionation factor per D, $10^3 \ln \alpha_{l-v}$, with the classical limit shown as a dotted line.
    Blue open squares, orange filled circles, and gray triangles denote PIMD, iso-EPIGS, and MD, respectively.
    All simulations employ the [MODEL] potential, with [P] beads in PIMD; details are given in Section~[X].
    Error bars in (c) represent [one standard error] and are smaller than the symbols; lines are guides to the eye.
    }
  \label{fig:hod}
\end{figure}

A central aim of iso-EPIGS is to predict isotopic compositions that never appear in training from models trained only on the fully hydrogenated and fully deuterated endpoints.
HDO is a clean and stringent test of this generalization: it places different masses on two hydrogen sites that are equivalent in every training molecule, and it is the form in which deuterium occurs in natural water, governing hydrogen-isotope fractionation in the water cycle.\cite{Craig1961,Dansgaard1964,Gat1996}
Thermodynamic quantities depend on the total quantum kinetic energy of the molecule, and these transfer well to HDO.
For properties set by the total quantum kinetic energy, such as $\Delta H_{\mathrm{vap}}$, iso-EPIGS describes HDO as accurately as the training endpoints: its kinetic energy agrees with PIMD to within 1~meV per molecule in both phases (SI), and its vaporization enthalpy is reproduced over the full temperature range (Fig.~\ref{fig:hod}).

Isotope fractionation depends on the kinetic energy of the substituted atom alone and is far more sensitive.
We compute the liquid--vapor fractionation factor $\alpha$ for $\mathrm{H_2O(l)+HDO(v)}\rightleftharpoons\mathrm{H_2O(v)+HDO(l)}$ from
\begin{equation}
10^3\ln\alpha_{\mathrm{l\text{-}v}}=-\frac{10^3}{k_{\mathrm B}T}\left(\Delta A^{\mathrm{H}\rightarrow\mathrm{D}}_{\mathrm l}-\Delta A^{\mathrm{H}\rightarrow\mathrm{D}}_{\mathrm v}\right),
\label{eq:alpha}
\end{equation}
where $\Delta A^{\mathrm{H}\rightarrow\mathrm{D}}_{\mathrm l}$ and $\Delta A^{\mathrm{H}\rightarrow\mathrm{D}}_{\mathrm v}$ are the free energies of replacing one hydrogen atom by deuterium in the liquid and in the vapor, obtained by Mass-TI (Eq.~\ref{eq:mass_ti}).\cite{Ceriotti_JChemPhys_2013_v138_p14112,MarklandBerne2012}
Classically $\alpha=1$, so the fractionation is a purely nuclear quantum effect, and a sensitive one: an error of 1~meV in the liquid--vapor difference of the integrand changes $10^3\ln\alpha$ by about 12.
PIMD gives $10^3\ln\alpha$ decreasing from $144.6\pm0.8$ at 281~K to $42.6\pm0.5$ at 409~K (Fig.~\ref{fig:hod}).
iso-EPIGS deviates from PIMD by only $+0.6$ to $+3.6$, and the kinetic energies of the substituted D and the remaining H in gas-phase HDO at 300~K deviate by $+1.3$ and $-1.4$~meV.
Deuterium is enriched in the liquid because the hydrogen nuclei are more strongly confined there, as the larger liquid kinetic-energy correction in Fig.~\ref{fig:water_K_hvap} shows, and the enrichment weakens as quantum effects decline with temperature.\cite{MarklandBerne2012}
The weak non-additivity of successive H$\rightarrow$D substitutions in water motivates a predominantly local description of the isotope response.\cite{Polyakov2005GeometricMean} 
Iso-EPIGS combines kinetic-energy supervision of individual mass derivatives with a bond-local mass readout (Eq.~\ref{eq:isotope_readout}). 
This representation limits reliance on correlations between hydrogen masses that always vary together in the endpoint training data. 
Removing the locality restriction increases deviations in $10^3\ln\alpha$ from $0.6$--$3.6$ to $8$--$16$ (SI), supporting the importance of this physical constraint. 
Trained only on $\mathrm{H_2O}$ and $\mathrm{D_2O}$, iso-EPIGS therefore resolves the fractionation thermodynamics of unseen HDO, demonstrating robust transferability without additional path-integral training data.

% \subsubsection*{Computational cost and scaling with isotopic composition}
\subsubsection{Isotope effects at classical cost}
\begin{figure}
    \centering
    \includegraphics[width=0.6\columnwidth]{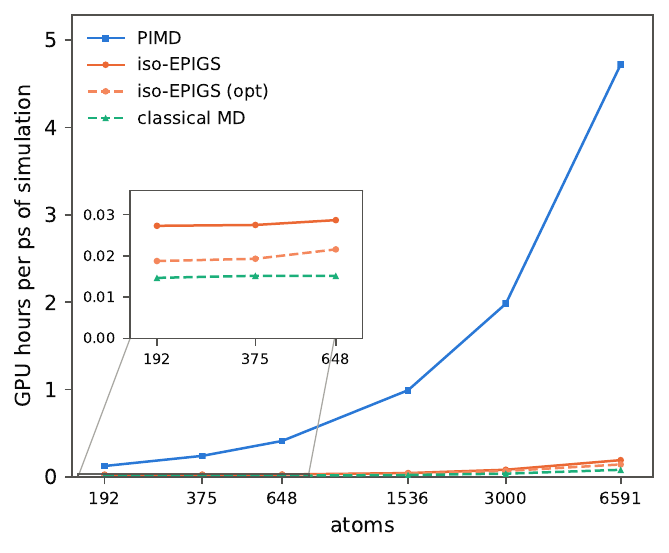}
    \caption{
        Computational cost per picosecond of simulated time on a single NVIDIA V100 (32~GB) GPU for periodic liquid water boxes of 64--2197 molecules at 300~K, all on the MACE-OFF24 potential.
        Shown are PIMD with 30 beads (time step 0.25~fs), iso-EPIGS centroid MD (MACE-OFF24 plus the learned correction, 0.5~fs, including the post-processing of $\partial\Delta A_{\mathrm c}/\partial g_i$), its optimized implementation (opt), and classical MD on MACE-OFF24 alone as a lower bound.
        The inset magnifies the three classical-cost methods for the three smallest boxes.
    }
    \label{fig:cost}
\end{figure}

Iso-EPIGS accelerates quantum-equilibrium sampling by propagating a single centroid rather than a full ring polymer.
We benchmarked liquid water at 300~K on a single NVIDIA V100 GPU, using MACE-OFF24 as the underlying potential (Fig.~\ref{fig:cost}).
PIMD used 30 beads with a 0.25~fs time step, whereas iso-EPIGS used a 0.5~fs time step, including evaluation of the learned free-energy correction and the mass derivatives required for mass-TI.
For 2197 water molecules, the cost decreased from 4.7 to 0.19~GPU hours per ps, corresponding to a 24-fold speedup.
The advantage is expected to increase at lower temperatures, where converged PIMD requires more beads while iso-EPIGS retains its single-centroid representation.

These timings exclude reference-data generation and training, whose costs are amortized over subsequent simulations.
Iso-EPIGS follows a ``learn small, predict big'' strategy: fixed-centroid reference calculations on monomers and small clusters yield models applicable to large condensed-phase systems.
This parallels the use of machine-learning potentials to amortize expensive ab initio calculations, extending the same strategy to path-integral sampling.
The resulting models can be reused across system sizes, temperatures, phases and isotopic compositions.
For isotope-substitution free energies, conventional mass-TI requires PIMD sampling at intermediate masses.
Iso-EPIGS instead learns the continuous mass dependence from fully hydrogenated and fully deuterated reference data alone, then performs mass-TI through near-classical centroid simulations.
Intermediate masses and unseen partially deuterated isotopologues therefore require no additional path-integral training data.

\section{Conclusions}

Learning the continuous, atom-resolved mass dependence of the centroid free energy yields a transferable description of isotope effects.
iso-EPIGS realizes this by conditioning EPIGS on per-atom masses, constraining its mass derivatives with centroid quantum kinetic energies, and localizing the mass dependence to each atom's own environment.
Benchmarks across molecular, liquid, and crystalline systems show that iso-EPIGS resolves isotope shifts in dimer binding and acid-dissociation free energies to within 1.2~meV of path-integral references, reproduces the quantum kinetic energies and vaporization enthalpies of water, recovers the PIMD sublimation enthalpies of oxalic acid within statistical uncertainty, and captures the isotope dependence of its lattice contraction.
These results cover both quantum weakening and strengthening of intermolecular cohesion across diverse hydrogen-bond environments.
Because the model provides both the free energy and its mass derivatives, isotope-substitution free energies can be obtained through two independent thermodynamic-integration routes, whose closure to about 0.1~meV offers a consistency check that requires no reference data.
Trained only on fully hydrogenated and fully deuterated systems, the model also transfers to mixed isotopologues, reproducing the liquid--vapor fractionation of HDO, which is sensitive to errors of 1~meV, so training effort need not scale with the number of isotopologues.
These capabilities enable quantitative predictions of equilibrium isotope fractionation in complex systems, supporting the design of isotope-separation materials and the interpretation of environmental isotope signatures.
More broadly, iso-EPIGS brings within reach foundation models of isotope thermodynamics spanning diverse chemical space.

\bibliography{ref}
\end{document}